\documentclass[aps, pra, twocolumn, reprint, final, showpacs, superscriptaddress, floatfix, amssymb, amsmath, amsthm, hidelinks, numbers, sort&compress, floats]{revtex4-1}

\usepackage{units}
\usepackage{graphicx}
\usepackage{float} 
\usepackage{upgreek}  %<-- Para não deixar as letras gregas em itálico. ao invez de \mu, escreve-se \upmu
\usepackage{bm}  % <-- allows you to write bold greek letters
\usepackage[latin1]{inputenc}
\usepackage{url}
\usepackage{times}
\usepackage{epsfig}
\usepackage{xcolor}
\usepackage[colorlinks=true,citecolor=blue,allcolors=blue]{hyperref}  % hyperlinks
\usepackage{ae,aecompl} % <-- High-quality PDF

\begin{document}

\title{Out-of-phase oscillation between superfluid and thermal components for a trapped Bose condensate under oscillatory excitation}

\author{P.~E.~S.~Tavares}\email{pedro.ernesto.tavares@usp.br}
\author{G.~D.~Telles}
\author{R.~F.~Shiozaki}
\author{C.~Castelo~Branco}
\author{K.~M.~Farias}
\author{V.~S.~Bagnato}\email{vander@ifsc.usp.br}
\affiliation{Instituto de F\'{\i}sica de S\~ao Carlos, Universidade de S\~ao Paulo, Caixa Postal 369, 13560-970 S\~ao Carlos, SP, Brazil}

\date{February 8, 2013}

\begin{abstract}
The vortex nucleation and the emergence of quantum turbulence induced by oscillating magnetic fields,  introduced by Henn~E~A~L, {\it et al.} 2009 ({\it Phys. Rev. A} \textbf{79} 043619) and Henn~E~A~L, {\it et al.} 2009 ({\it Phys. Rev. Lett.} \textbf{103} 045301), left a few open questions concerning the basic mechanisms causing those interesting phenomema. Here, we report the experimental observation of the slosh dynamics of a magnetically trapped $^{87}$Rb Bose-Einstein condensate (BEC) under the influence of a time-varying magnetic field. We observed a clear relative displacement in between the condensed and the thermal fraction center-of-mass. We have identified this relative counter move as an out-of-phase oscillation mode, which is able to produce ripples on the condensed/thermal fractions interface. The out-of-phase mode can be included as a possible mechanism involved in the vortex nucleation and further evolution when excited by time dependent magnetic fields.
\end{abstract}

\maketitle

\section{Introduction}

	The research evolution involving atomic superfluid during the last decade has been quite significant. It started with the demonstration of the superfluidity associated with Bose-Einstein condensation (BEC) \cite{Marago2000}, the nucleation of vortices \cite{Matthews1999}, the formation of vortex lattices \cite{Madison2000}, and finally the first observation of quantum turbulence in a $^{87}$Rb BEC \cite{Henn2009a}. Such evolution presents the full potential of using atomic superfluid samples as workbenches for many phenomena that could only be experimentally accessed before using liquid helium \cite{Allen1938}. Recently, an oblate sample of trapped atoms, corresponding to a 2D-superfluid, has been explored in many aspects related to superfluidity. A repulsive obstacle passing through this 2D system has generated vortex-antivortex pairs and its rotation dynamics could be studied \cite{Neely2010}. In the meanwhile, theory has also progressed in many directions \cite{Kobayashi2007,Nowak2011,Berloff2010}. In a recent experiment on generation of vortex by applying an oscillatory field in a trapped Bose-Einstein condensate \cite{Henn2009}, the proliferation of vortices with amplitude and time of excitation \cite{Seman2010,Seman2011} results in a configuration of vortex filaments in many spatial directions, allows to obtain a turbulent cloud of atomic superfluid \cite{Henn2009a}. In those already reported experiments, the mechanisms from which vortices were generated was still an open question. Explanation were based on a few theoretical simulation \cite{Parker2005}. Experimental evidences of nucleation of vortices in oscillatory conditions, previously to vortices proliferation and turbulence, was a missing fact in the sequence of experiments. This report brings the first evidences of excitations that can eventually promote vortex nucleation during oscillatory excitation. 
	
	To understand our present experiment we shall start with a widely explored scenario in liquid helium, below the $\lambda$-point, which corresponds to the two-fluid system created by the coexistence of a superfluid and a normal fluid fraction \cite{Blaauwgeers2002}. In this case, the normal fluid can be regarded as a classical viscous fluid. Experiments in which a counterflow between the two parts became possible, resulted in many important experimental investigations including the equivalence of classical flow using the Reynolds number criterion \cite{Finne2003} and the generation of vortices.
	
	The two-fluid system seems appropriate to help answering questions concerning the onset of dissipation on the superfluid flow. One of most interesting counterflow experiments is the thermal counterflow \cite{Finne2003}. In that case, the temperature gradient generates a superfluid flow opposite to the normal fluid flow, giving rise to a mutual frictional force between the two fluids \cite{Gorter1949}. Although thermal counterflow was not yet been realized in a sample of trapped atomic superfluid, a collective mode presenting an out-of-phase oscillation between the condensate and thermal cloud has been investigated systematically by references \cite{Stamper-Kurn1998,Meppelink2009}.
	
	In this report we show that, under the influence of a oscillatory magnetic field, a trapped BEC may evolve to a scenario where the condensed and the thermal fractions present a relative out-of-phase motion, characterizing the onset of counterflow (second sound type excitation). Such situation may be useful to investigate the circumstances in which quantized vortices and turbulence take place during oscillatory excitation.
	
\section{Experimental setup and time sequence}

	The experimental sequence used to produce the BEC and nucleate vortices in our condensate can be summarized as follows. First, a $^{87}\rm{Rb}$ BEC containing about $1 \times 10^{5}$ atoms with a small thermal fraction is produced in a cigar-shaped QUIC magnetic trap with: $\nu_{r}=\unit[210]{Hz}$, and $\nu_{z}=\unit[23]{Hz}$ trapping frequencies. While the BEC sample is realized and still held inside the trap, a weak, sinusoidal, time-varying magnetic field is turned on. This field is superposed to the QUIC trap and is generated by an extra pair of Helmholtz coils mounted with the symmetry axis aligned in a small angle relative to the Ioffe coil axis. Since the coils are axially misaligned, the gradient near the trapping region present field components parallel to each of the trap eigen axis. This time-varying approach was originally developed to investigate coherent mode excitations in our BEC samples.
	
	A combination of shape modifications, trap minimum displacement and axial twisting couples to the atoms during the excitation. The perturbing field was on for time lapses usually ranging from $20$ to $\unit[100]{ms}$; and it oscillates sinusoidally in time with frequencies in between $150$ to $\unit[210]{Hz}$. There is also an offset chosen so that during the oscillation period it goes from zero to the maximum amplitude and then back to zero, always pushing and never pulling. The maximum amplitude reached by the field gradient characterizes the perturbing strength and has been varied from zero to $\unit[190]{mG/cm}$ along the vertical axis.
	
	Finally, the atoms are held trapped for another $\unit[20]{ms}$, after the time-varying field is turned off. The measurements are performed via time-of-flight absorption images, with the atomic cloud is expanding in free fall. For small amplitudes and short time lapses we observe dipolar, quadrupolar, and scissor modes but no vortices. By increasing the values of both parameters we were able to produce vortices, which grew in number with the amplitude and/or the excitation elapsed time. Further information can be found elsewhere \cite{Henn2008,Henn2009a,Henn2009,Henn2009b}.
	
	During the experimental runs, the clouds may undergo a few different excitation modes before the vortices are produced. To better observe the modes we kept the current in the coils oscillating around $\unit[170]{Hz}$ during $t_{exc}=\unit[100]{ms}$. Then, we wait a time $\tau$ lapse before releasing the sample from the trap, allowing it to freely expand for $\unit[23]{ms}$. A resonant probe laser pulse of $\unit[70]{\mu s}$ takes an absorption image. The RF forced evaporation was adjusted to keep the condensed fraction at about $\unit[60]{ms}$. The temperature of the cloud was kept around $\unit[100]{nK}$, for the conditions of this experiment. After the absorption image is acquired, we fit the images using a bimodal profile, constituted by a Thomas-Fermi and a Gaussian profiles for the condensed and the thermal fractions. Finally, different aspects of these two components are observed as a function of the hold time, $\tau$.
	
\section{Overall motion of the cloud during oscillation}

\begin{figure}[tbp]
			\centering
			\includegraphics[width=3cm]{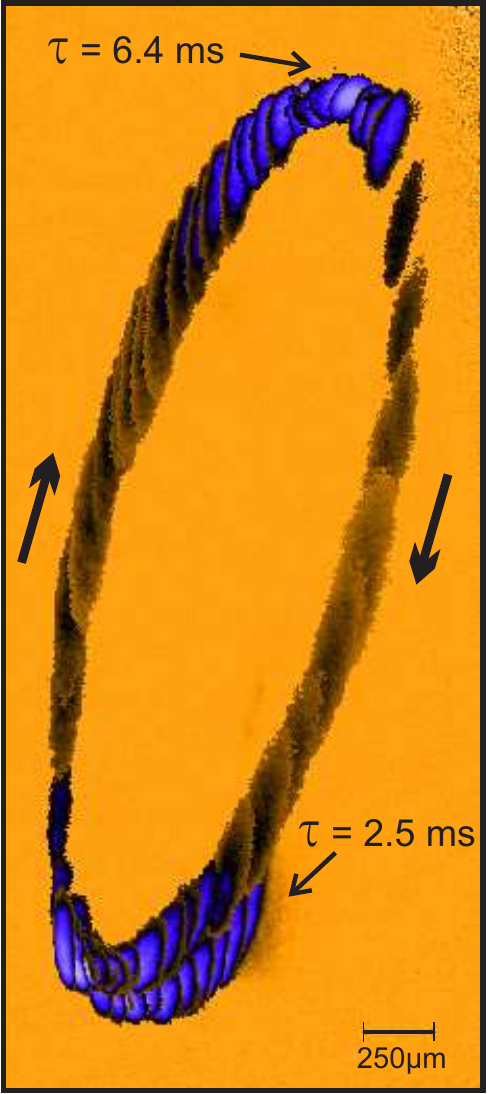}
			\caption{\label{Fig1} Overview of the cloud's motion path after $\unit[23]{ms}$ of time-of-flight (TOF) absorption images. The arrows point towards the increasing hold time, $\tau$.}
		\end{figure}

	To follow the cloud's motion into the trap, we have taken a sequence of TOF snapshots varying $\tau$. The superposed images were stacked together to better visualize the full path, as shown in figure~\ref{Fig1}. The arrows point towards the increasing $\tau$, evolving clockwise. As the excitation progresses, a variety of collective modes take place absorbing the energy pumped into the cloud by the oscillating magnetic field.
	
	First, the clouds' center-of-mass oscillate inside the trap in the dipole mode \cite{Jin1996}. This mode corresponds to the full oscillation, in trap, with the trap characteristic frequency in each direction. Besides the dipole mode, the long and short axes of the cloud also oscillate out-of-phase, characterizing the collective quadrupole mode of excitation \cite{Mewes1996a}. Since the longitudinal and transversal directions do not oscillate exactly $90^{\circ}$ out-of-phase, we believe that higher order modes might also be present.
	
	Additionally to these collective modes, the tilting of the condensed cloud's long axis is also observed as previously reported \cite{Henn2009}. This tilting oscillation is related to scissors mode \cite{Marago2000} and shows that the excitation is able to produce rotation in the cloud.

	An interesting observation seen in figure~\ref{Fig1} shows that the dipole mode trajectory does not close on itself which is an evidence that the dipole mode is also excited in the direction of the weak confinement (lower frequency). In fact, the cloud is under dipole excitation in all three directions, with different amplitudes, resulting in a more complex trajectory for this cyclic path.

\begin{figure}[tbp]
			\centering
			\includegraphics[width=3.9cm]{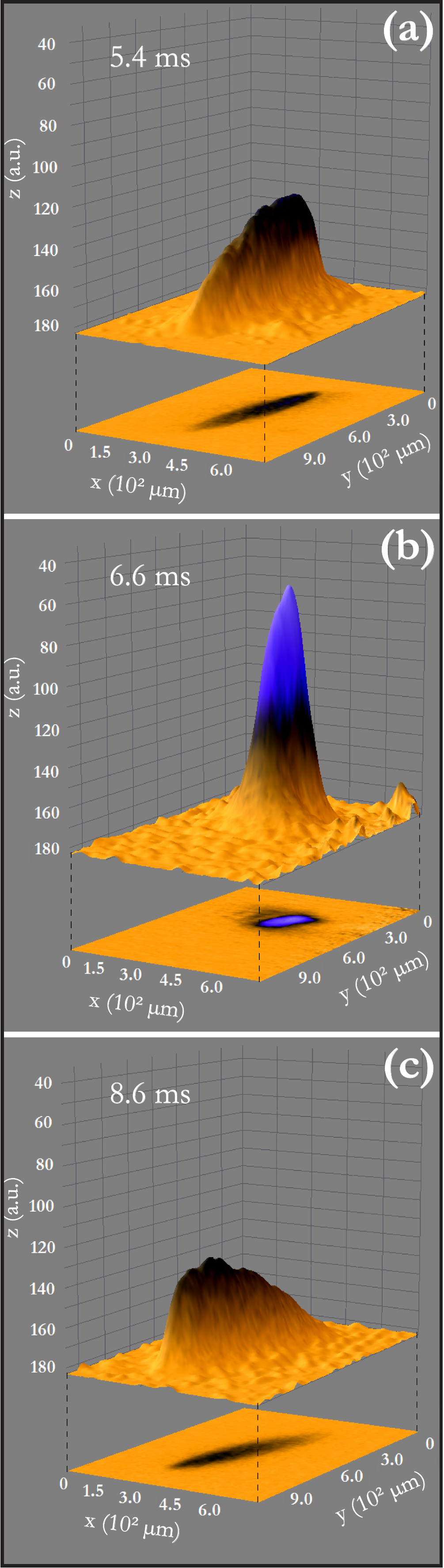}
			\caption{\label{Fig2} Sequence of typical picked density profiles showing the relative motion in between the thermal (normal fluid) and the condensed (superfluid) fractions.}
		\end{figure}

	The excitation of the collective modes seems to play an important role on nucleating vortices in our system \cite{Henn2009}. By changing the excitation frequency we were able to vary the amplitude of oscillation of those described modes. We also observed a relative motion between the two components of the fluid. This out-of-phase oscillation in between the thermal cloud and the condensate correspond to a mode previously investigated in detail by Ketterle and van der Straten \cite{Stamper-Kurn1998,Meppelink2009}. It is quite surprising that this mode is coupled to the dipole mode and it may be fundamental to the vortices nucleation. The condensed and the thermal fractions oscillate with respect to each other in a macroscopic scale such that there is a relative motion between the two centers of mass. Figure~\ref{Fig2} presents three typical density profiles during the oscillation. The presence of such mode, excited by the oscillatory motion, is quite relevant to understand the mechanisms of vortex generation and evolution \cite{Henn2009}.
	
\begin{figure}[tbp]
			\centering
			\includegraphics[width=7.5cm]{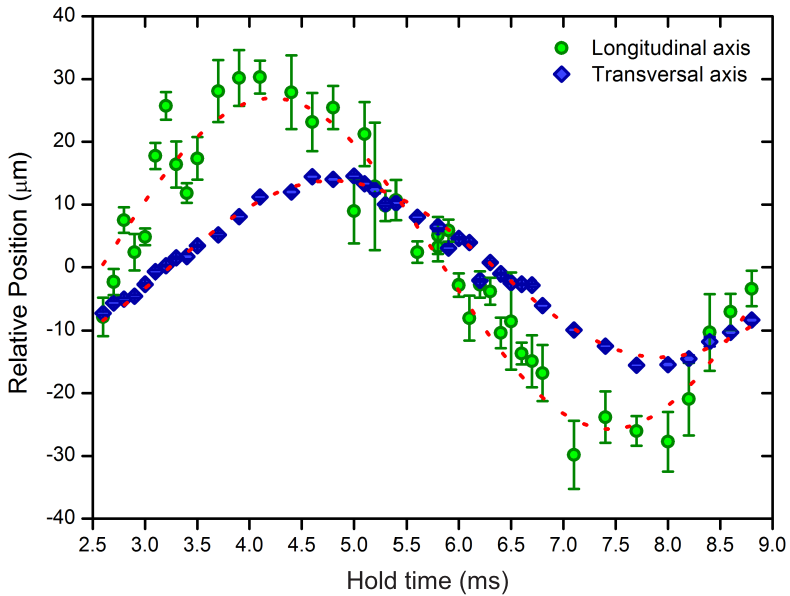}
			\caption{\label{Fig3} Relative position between the centers of mass of the superfluid and thermal components for the longitudinal and transversal directions as a function of hold time after the excitation. The dotted lines are sinusoidal fittings.}
		\end{figure}

	The largest amplitude for the relative motion takes place along the cloud longest axis, but it is also present along the short axis since the excitation is not limited to a single axis. To determine the \textit{in situ} sizes from the expanded cloud values, we calculated the scaling factors following the method developed  in Ref.~\cite{Castin1996}. The results for the \textit{in situ} counterflow motion are shown in figure~\ref{Fig3} for the longitudinal and the transversal directions. The smallest displacement observed between the two clouds corresponds to the upper and lower turning points seen in figure~\ref{Fig1}. In these points the dipole mode has the maximum acceleration and the counterflow has the maximum velocity.
		
	While the longitudinal relative position in between the two clouds center of mass reaches distances of about $\unit[30]{\mu m}$, the transverse relative motion reaches the largest displacement of about $\unit[12]{\mu m}$. Both motion occur in same frequency, as one can see on the plot in figure~\ref{Fig3}.
	
	We analyzed the motion of this out-of-phase mode using the force per unit volume as derived by Gorter and Mellink \cite{Gorter1949}:
	\begin{equation}\label{eq:Fsn}
		\mathbf{F}_{sn} = -A\rho_{s}\rho_{n}(\mathrm{v}_{ns}-\mathrm{v}_0)^{2}\mathbf{\mathrm{v}}_{ns} \ \ ,
	\end{equation}
where $\rho_{s}$ and $\rho_{n}$ are the densities of the superfluid and the normal fluid, respectively, $\mathrm{v}_0$ is a constant on the order of $\unit[1]{cm/s}$ and $A$ is a temperature dependent constant. It should be noted that Equation~(\ref{eq:Fsn}) is a phenomenological relation, which is proportional to the cubic of the relative velocity ($\mathbf{\mathrm{v}}_{ns} = \mathbf{\mathrm{v}}_{n} - \mathbf{\mathrm{v}}_{s}$) in between the two fluids, and it seems to be valid even beyond the turbulence onset in the superfluid. The origin of this interaction is the scattering of thermal excitation as discussed by W. F. Vinen \cite{Vinen1957}.
	
	The out-of-phase oscillatory motion corresponds to relative velocities up to $\mathrm{v}_{ns}\approx\unit[3]{cm/s}$ under our experimental conditions. This velocity can be larger or smaller, depending on the amplitude of the oscillating external magnetic field.
	
	To better understand the net result of the superfluid fraction slosh motion, we analyzed the surface ripples appearing on the boundary region separating the condensed and thermal fraction (see figure~\ref{Fig5}). The ripples may be described as undulations, or ruffling of the BEC/thermal in between surface giving it a wavy form and which may end up folded on itself like wave crests on the shore forming tubes or vortices. To analyze it, we have first fitted the thermal fraction only and then subtracted it from the image, resulting in a pure BEC shot, which was then fitted by the Thomas-Fermi profile and, finally, it also subtracted from the image, this resulted in a perimeter line surrounding the condensed fraction. 
	
\begin{figure}[tbp]
			\centering
			\includegraphics[width=7.3cm]{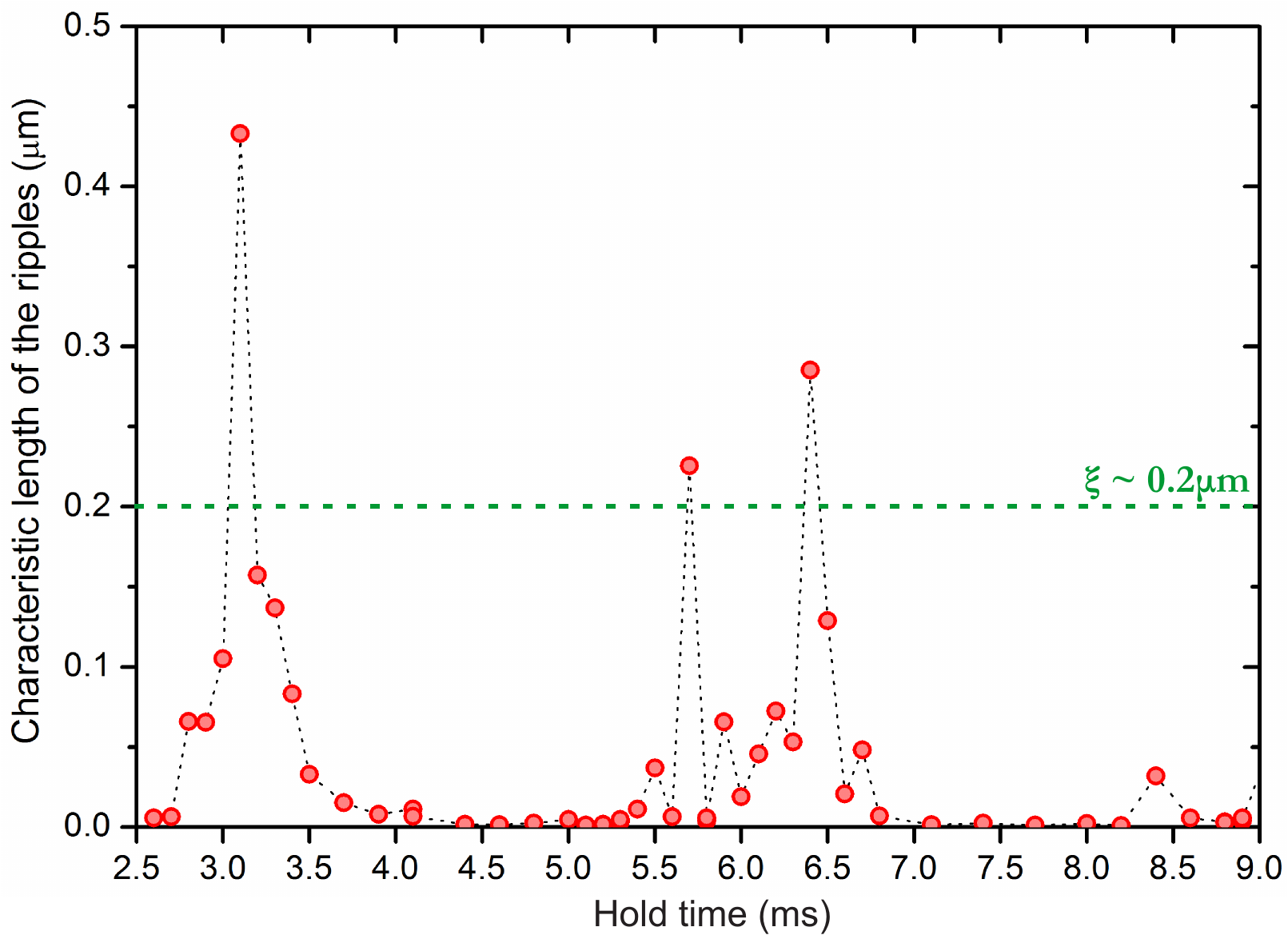}
			\caption{\label{Fig4} Characteristic length of the ripples in the superfluid surface as a function of the hold time. The dotted line is an eye guide only. For reference, we displayed the value of healing length.}
		\end{figure}

\begin{figure}[tbp]
			\centering
			\includegraphics[width=3cm]{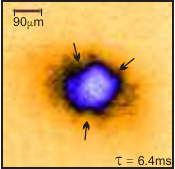}
			\caption{\label{Fig5} Typical time-of-flight absorption image showing the ripples on the superfluid surface see the black perimeter region due to the counterflow.}
		\end{figure}

	To help on the data analysis we defined the surface ripples characteristic length. It is determined as the highest peak resulting from the Fourier transform over the perimeter line separating the condensed and the thermal fractions on the sloshed clouds. In figure~\ref{Fig4}, we show the characteristic length of the ripples versus the hold time in order to allow for comparing to figure~\ref{Fig3}. We observe that the ripples amplitude are much larger around the turning points, where the maximum relative velocity in between the superfluid and normal fraction takes place.
			
	The larger the ripples the more likely to nucleate vortices into the superfluid fraction it will be during the slosh motion. To show these surface ripples, we present a typical time-of-flight cloud in figure~\ref{Fig5}. As it can be seen, the ripples may be the starting the vortex nucleation and one might be able to see the characteristic shape of the surface ripples already start to looking like vortices. In fact, a closer look reveals the vortex nucleation around the regions where the fluctuations are larger.
	
	The experiment may be understood as a counterflow of the superfluid over the normal fraction. We believe that when the counterflow exceeds the critical velocity, the surface ripples grow larger, and may induce the formation of vortices. In our experiment due to the oscillatory nature of the counterflow, the maximum velocity is sustained for a short time lapse, with small vortex nucleation less likely evolving to the turbulent regime. Different experimental condition with larger amplitudes already demonstrated the onset of turbulence \cite{Henn2009a}.
		
	Assuming the mutual friction force, Equation~(\ref{eq:Fsn}), as the cause for the increasing fluctuations on the interface between the two fractions, the regions with larger ripples correspond to $\mathrm{v}_{ns} > \mathrm{v}_0$. Looking at figure~\ref{Fig4}, for $\tau$ running from $3.5$ to $\unit[5.5]{ms}$, $\mathrm{v}_{ns} < \mathrm{v}_0$ and fluctuations are kept low (basically zero characteristic length). Around $\unit[5.5]{ms}$, the relative velocity must be close to $\mathrm{v}_0$, and from there on the increase in $\mathrm{v}_{ns}$ shall exceed $\mathrm{v}_0$. Hence, we estimate $\mathrm{v}_0$ to be about $\unit[2.5]{cm/s}$.
	
	We have analyzed the situation using the arguments found in \cite{Finne2003}, where the turbulence in superfluids is investigated in terms of two dimensionless parameters: $q$ and $Re_s$. Basically, the intrinsic parameter $q$ deals with the relative importance of two competing terms determining the energy dissipation in the system. The extrinsic parameter $Re_s$, the superfluid Reynolds number, which characterizes the flow of the superfluid fraction, given by \cite{Finne2003}:
	\begin{equation}\label{eq:Res}
		Re_{s} = \frac{m\xi v}{2\pi\hbar} \ \ ,
	\end{equation}
where $\xi$ is the healing length, $v$ the flow velocity and $m$ is the atomic mass.

	These parameters can be considered in the case of two mutually penetrating fluid components. Considering $v \equiv \mathrm{v} _{ns}$ and $\xi = (8\pi\rho_{s}a_s)^{-1/2}$, where $a_s$ is the s-wave scattering length ($\sim100a_0$) and $\rho_{s}=\unit[2.5\times10^{14}]{atoms/cm^3}$, we obtain $Re_s$ running from $0$ to $1$. According to reference \cite{Finne2003}, these values are not large enough to generate turbulence. This statement is in agreement with our observations, where the fluctuations are able to nucleate vortices but they do not evolve to a turbulent state. The normal component of the fluid vanishes close to $T=\unit[0]{K}$ stopping the counterflow and the frictional force. As a result, we should expect the vortex nucleation and the evolution of the vortex configuration to become more difficult as the temperature goes to zero. In fact, this justifies some of the previous observations \cite{Shiozaki2011} as well as the need to introduce dissipation in the theoretical models to be able to predict the turbulence emergence \cite{Seman2011}. 
	
\section{Conclusions} 

	We have investigated the slosh move of a non pure BEC sample, containing a condensed (superfluid) embedded in a thermal fraction (normal fluid), excited by time-varying magnetic fields. The relative motion between the two components was interpreted as an out-of-phase oscillation mode, previously investigated by other groups \cite{Stamper-Kurn1998,Meppelink2009}. It was observed that ripples are more intense near the time  instants of largest relative velocity, where high amplitude ripples are generated on the superfluid/normal fluid interface, creating the right conditions for the vortex nucleation. Our findings helped on explaining previous experimental results \cite{Henn2009}, as well as the presented results may bring new possibilities for exploring the superfluid character of atomic BECs.
	
%%%%%%%%%%%%%%%%%%%%%%%%%%%%%%

\begin{acknowledgments}
This work was supported by Funda\c{c}\~ao de Amparo \`{a} Pesquisa do Estado de S\~ao Paulo, Conselho Nacional de Desenvolvimento Cient\'{i}fico e Tecnol\'{o}gico, and Coordena\c{c}\~ao de Aperfei\c{c}oamento de Pessoal de N\'{i}vel Superior. We thank the collaboration of D. V. Magalh\~aes, J. Seman, K. Merloti, G. G. Bagnato, A. V. M. Marino, M. Martinelli and C. J. Villas-B\^oas for technical support and fruitful discussions.
\end{acknowledgments}

%%% ==> BIBLIOGRAPHY:
\bibliographystyle{apsrev4-1}
\bibliography{outofphaseReferences}

\end{document}